\documentclass{emulateapj} 
\begin{document}

\title{The Megamaser Cosmology Project. III. Accurate Masses of Seven
Supermassive Black Holes in Active Galaxies with Circumnuclear Megamaser
Disks}                                                                  
 
\author{C. Y. Kuo\altaffilmark{1}, J. A. Braatz\altaffilmark{2},
J. J. Condon\altaffilmark{2}, C. M. V. Impellizzeri\altaffilmark{2}, K.Y. Lo\altaffilmark{2},
I. Zaw\altaffilmark{6},   M. Schenker\altaffilmark{5},
C. Henkel\altaffilmark{4}, M. J. Reid\altaffilmark{3}, J. E. Greene\altaffilmark{7}}
 
\affil{\altaffilmark{1}Department of Astronomy, University of Virginia,
Charlottesville, VA 22904}                                              
\affil{\altaffilmark{2}National Radio Astronomy Observatory, 520
Edgemont Road, Charlottesville, VA 22903, USA}                          
\affil{\altaffilmark{3}Harvard-Smithsonian Center for Astrophysics,
60 Garden Street, Cambridge, MA 02138, USA}                             
\affil{\altaffilmark{4}Max-Planck-Institut f\"ur Radioastronomie,
Auf dem H\"ugel 69, 53121 Bonn, Germany}                                
\affil{\altaffilmark{5}Department of Astronomy, California Institute
of Technology, CA 91125 }                                               
\affil{\altaffilmark{6}New York University Abu Dhabi, Abu Dhabi, UAE } 
\affil{\altaffilmark{7}Department of Astrophysical Sciences, Princeton
University, Princeton, NJ 08544, USA}

\begin{abstract} 
Observations of H$_2$O masers from circumnuclear disks in active galaxies for the
Megamaser Cosmology Project allow accurate measurement of the mass of
supermassive black holes (BH) in these galaxies. We present the
Very Long Baseline Interferometry (VLBI) images and kinematics of
water maser emission in six active galaxies: NGC~1194, NGC~2273,
NGC~2960 (Mrk~1419), NGC~4388, NGC~6264 and NGC~6323.  We use the
Keplerian rotation curves of these six megamaser galaxies, plus a
seventh previously published, to determine accurate enclosed masses
within the central $\sim0.3$ pc of these galaxies, smaller
than the radius of the sphere of influence of the central mass in all cases.
We also set lower limits to the central mass densities of between 0.12 and 60 $\times
10^{10} M_{\odot}$~pc$^{-3}$.  For six of the seven disks, the high
central densities rule out clusters of stars or stellar remnants as
the central objects, and this result further supports our assumption
that the enclosed mass can be attributed predominantly to a supermassive
black hole.  The seven BHs have masses ranging between 0.76
and 6.5$\times$10$^7 M_{\odot}$.  The BH mass errors are
$\approx11$\%, dominated by the uncertainty of the Hubble constant. 
We compare the megamaser BH mass determination with other
BH mass measurement techniques.  The BH mass based on virial
estimation in four galaxies is consistent with the megamaser
BH mass given the latest empirical value of $\langle f \rangle$, but the virial mass
uncertainty is much greater.  Circumnuclear megamaser disks allow the best mass
determination of the central BH mass in external galaxies and significantly improve
the observational basis at the low-mass end of the $M-\sigma_{\star}$ relation.  
The $M-\sigma_{\star}$ relation may not be a single, low-scatter power law
as originally proposed. MCP observations continue and we expect
to obtain more maser BH masses in the future.
                                                    
\end{abstract} 
 
\keywords{accretion, accretion disks --
galaxies: nuclei -- galaxies: masers -- galaxies: active --
galaxies: ISM -- galaxies: Seyfert}

\section{INTRODUCTION} 
The primary goal of the Megamaser Cosmology Project (MCP; Braatz et
al. 2009; Reid et al.  2009a; Braatz et al. 2010) is to determine the
Hubble constant $H_0$ to $\sim$ 3$\%$ accuracy in order to constrain
the equation of state parameter $w$ of dark energy.  The key to
achieving this goal is to measure accurate distances to galaxies well
into the Hubble flow (50 $-$ 200\,Mpc).  A proven method to measure
accurate angular-diameter distances involves sub-milliarcsecond
resolution imaging of H$_{2}$O maser emission from sub-parsec
circumnuclear disks at the center of active galaxies, a technique
established by the study of NGC 4258 with VLBI \citep{hmg99}. Throughout the
paper, we will refer to such circumnuclear disks as ``megamaser
disks'' and galaxies that contain such megamaser disks as ``megamaser
galaxies''.  As the MCP discovers and images megamaser disks, an
important result, in addition to the distance determination, is the
accurate measurement of the enclosed masses of the massive dark
objects (MDOs; Kormendy \& Richstone 1995; Richstone et al. 1998) at
the centers of these megamaser galaxies. We assume that the MDOs are
black holes (BHs) and justify this assumption in section 4.

The VLBI plays a crucial role in measuring BH masses with high precision especially 
in galaxies with lighter supermassive black holes (i.e. $M_\bullet \sim 10^{6}-10^{7} M_{\odot}$ ). 
The critical advantage provided by VLBI is angular resolution two
orders of magnitude higher than the best optical resolution.  For any
given galaxy with a nearly constant central mass density of stars, $M_{\rm BH}$
 is proportional to $R_{\rm inf}^3$, where $R_{\rm inf}$ is the radius of
the gravitational sphere of influence of BH \citep{barth03}. So, a factor of
100 increase in resolution permits measurements of masses up to $10^6$
times smaller.  Similarly, the central density limits that can be set
are up to $10^6$ times higher, high enough to rule out extremely dense
star clusters as the MDOs based on dynamical argument (see section 4) 
in most megamaser disks presented here.
 
Megamaser disks, such as the archetypal one in NGC 4258, are usually
found in the centers of Seyfert 2 galaxies. Because maser emission is
beamed and long path lengths are required for strong maser
amplification, the megamaser disks are observable only if the the disk
is close to edge-on. In NGC 4258 the disk is inclined $\sim$82$^\circ$
and its rotation curve is Keplerian to better than 1\% accuracy, which
makes the BH mass determination robust with very few assumptions. The
megamaser disks are typically smaller ($r\sim$0.2 pc in NGC 4258) than
the gravitational sphere of influence of their supermassive black holes ($r \sim 1$ pc in
NGC 4258); this guarantees that the gravitational potential is dominated by
the central mass.  Applying dynamical arguments to the 
extremely high mass density $\rho_0 \sim
10^{11} M_\sun$\,pc$^{-3}$ within the NGC 4258 maser disk
can rule out a dense cluster of
stars or stellar remnants for most of the central dark mass
\citep{mao95}, implying that the measured central mass is dominated by
a BH.

The megamaser disk method for estimating BH masses has some practical
limitations. First, finding megamaser disks is difficult, partly because
detectable disks need to be nearly edge-on.
Only eight BH masses have been published based on measurements of
megamasers : NGC~1068 \citep{ggr96}, NGC~2960 (Mrk~1419) (Henkel et al. 2002; not based on VLBI),
NGC~3079 \citep{kgm05}, NGC~3393 \citep{kgm08}, UGC~3789 (Reid et al. 2009; MCP paper
I), NGC~4258 \citep{hmg99}, NGC~4945 \citep{gmh97} , and Circinus
\citep{gbe03}. Second, some rotation curves are
significantly flatter than Keplerian, e.g. NGC 1068; NGC
3079; and IC 1481 \citep{mam09}. The origin of the flatter rotation curves
in these galaxies is unclear. It could be caused by self-gravity of a
massive disk, the presence of a nuclear cluster, or radiation pressure
\citep{lb03}. In these cases, the measured enclosed mass of the
disk could be significantly different from the actual BH
mass. Therefore, the key to measuring reliable BH masses using
megamasers is to find megamaser disks that have well defined edge-on
disks with Keplerian rotation curves.

Over the past two decades, there has been substantial progress in
detecting BHs and constraining their masses, especially in quiescent
or mildly active nearby galaxies \citep{kr95, kg01,kmd04,fer05}.  This
rapid progress was mainly facilitated by the high angular resolution
provided by the Hubble Space Telescope (HST) and by gradually maturing
techniques for modeling stellar dynamics of galaxies. The number of BH
detections has increased to the degree that the field has shifted from
confirming the existence of BHs to comparing their properties with
those of the host galaxies. One of the most discussed relationships
found between BHs and host galaxies is the correlation between BH mass
and the velocity dispersion of stars in the bulge (ie, the
$M_{\rm BH}$-$\sigma_{\star}$ relation; Ferrarese \& Merritt (2000);
Gebhardt et al. (2000); G\"{u}tekin et al. (2009), and references
therein). It is often suggested that this relation is a manifestation
of a causal connection between the formation and evolution of the
black hole and the bulge.  As pointed out by \citet{gbr00}, it is
natural to assume that bulges, black holes, and quasars formed, grew,
or ``turned on'' as parts of the same process, in part because the
collapse or merger of bulges might provide a rich fuel supply to a
centrally located black hole.  However, the nature of this connection
is still not well understood.
 
Clearly, accurate BH mass measurements in megamaser galaxies can help
to improve our understanding of the $M_{\rm BH}$-$\sigma_{\star}$
correlation, especially at the low-mass end.  We present
high-resolution images and rotation curves of megamaser disks in six
new galaxies and their BH masses.  In addition, we re-analyze the BH
mass in UGC 3789 using the data from \citet{rbg09} and present the new
measurement here. In section 2, we present our sample of galaxies,
VLBI observations and data reduction. Section 3 shows the VLBI images
and rotation curves of the megamaser disks, followed by our analysis
of BH masses. In section 4, we rule out compact clusters of
stars or stellar remnants as the central dominant masses for the majority of 
the megamaser disks. In section 5, we compare results from other mass measuring techniques 
to our maser masses. The discussion of our main results is presented in 
section 6, and we summarize the results in section 7. A more in-depth discussion of the
$M_{\rm BH}$-$\sigma_{\star}$ relation, including the new maser BH masses,
is presented in a companion paper by
\citet{gre10}.
 
\section{The Sample, Observation, and Data Reduction} 
\subsection{The Megamaser Disk Sample} 
Water masers have so far been detected in $\sim$ 136
galaxies.\footnote{https://safe.nrao.edu/wiki/bin/view/Main/MegamaserCosmologyProject}
Most of the megamasers originate in AGN \citep{lo05} and about 20\% of
them show spectra suggestive of emission from sub-parsec scale,
edge-on, circumnuclear disks. In the MCP, we are currently conducting
VLBI observations of six megamaser disks to determine the angular
diameter distances to their host galaxies.  While more data are needed
to measure their distances accurately, the quality of the VLBI imaging
is already good enough to measure precise central BH masses.  In
addition to the six galaxies for distance measurements, we also have
VLBI data on the megamaser disk in NGC 4388 to measure the central BH
mass, even though the data are not suitable for a distance
determination.  Table 1 lists coordinates, recession
velocities, spectral types, and morphological types for these seven
galaxies.
 
\subsection{Observations} 
The megamaser galaxies in our sample were observed between 2005 and
2009 with the Very Long Baseline Array (VLBA)\footnote{The VLBA is a
facility of the National Radio Astronomy Observatory, which is
operated by the Associated Universities, Inc. under a cooperative
agreement with the National Science Foundation (NSF).}, augmented by
the 100-m Green Bank Telescope (GBT) and in most cases the Effelsberg
100-m telescope.\footnote{The Effelsberg 100-m telescope is a facility
of the Max-Planck-Institut f\"ur Radioastronomie} Table 2 shows the
basic observing information including experiment code, date observed,
antennas used, and sensitivity.
 
We observed the megamasers either in a phase-referencing or
self-calibration mode. With phase-referencing we perform rapid
switching of the telescope pointing
between the target source and a nearby ($< 1^\circ$) phase calibrator
(every $\sim$ 50 seconds) to correct phase variations caused by the atmosphere.  In a
self-calibration observation, we use the brightest maser line(s) to
calibrate the atmospheric phase.  In both types of observations, we
placed ``geodetic'' blocks at the beginning and end of the
observations to solve for atmosphere and clock delay residuals for
each antenna
\citep{reid09}.  For NGC 6323 and NGC 6264, we also placed a geodetic
block in the middle of the observations to avoid the zenith transit
problem at the GBT and to obtain better calibration.  In each geodetic
block, we observed 12 to 15 compact radio quasars that cover a wide
range of zenith angles, and we measured the antenna zenith delay
residuals to $\sim1$ cm accuracy.  These geodetic data were taken in
left circular polarization with eight 16-MHz bands that spanned $\sim$
370 - 490 MHz bandwidth centered at a frequency around 22 GHz; the
bands were spaced in a ``minimum redundancy'' manner to sample, as
uniformly as possible, all frequency differences between IF bands in
order to minimize ambiguity in the delay solution \citep{ak09}. We also observed strong compact radio quasars
every $\sim$ 20 minutes to 2 hours in order to monitor the
single-band delays and electronic phase differences among and across
the IF bands.  The errors of the single-band delays are $<$ 1
nanosecond.

\subsection{Data Reduction} 
We calibrated all the data using the NRAO Astronomical Image
Processing System (AIPS).  Since the geodetic data and the maser data
used different frequency settings, we reduced them separately. For the
geodetic data, we first calibrated the ionospheric delays using total
electron content measurements \citep{wc00} and the Earth Orientation
Parameters (EOPs) in the VLBA correlators with the EOP estimates
from the US Naval Observatory
(http://gemini.gsfc.nasa.gov/solve\_save/usno\_finals.erp).  We
performed ``fringe fitting'' to determine the phases, single-band
delays, and fringe rates of IF bands of each antenna for every geodetic
source, and the multi-band delay of each antenna was determined from these solutions. Finally, we determined the residual tropospheric
delay and clock errors for all antennas using the multi-band delays. We
applied these corrections to the maser data as described in the next
paragraph. In five datasets for NGC 6323, we also made
antenna position corrections before the fringe-fitting process when the
positions used in the observation deviated from the latest USNO
(http://rorf.usno.navy.mil/solutions/) solutions by more than $\sim$ 2
cm.
 
For the maser data, after the initial editing of bad data, we
corrected for ionospheric delay and the EOPs
in the same way as for the geodetic dataset. We then corrected the sampler
bias in the 2-bit correlator. The amplitude calibration was done with the
information in the gain table and the system temperature table. We
corrected the interferometer delays and phases caused by the effects
of diurnal feed rotation (parallactic angle), and applied the
tropospheric delay and clock corrections obtained from the geodetic
data afterwards. The next step was to perform fringe fitting on one or
two scans of the delay calibrators to calibrate the electronic phase
offsets among and across IF bands. The frequency axes of the maser
interferometer spectra were then shifted to compensate for
the changes in source Doppler shifts over the observing tracks.
 
The final step in calibration was to solve for the atmospheric phase
variation by using either phase-referencing or self-calibration.  In
the phase-referencing mode, we ran the AIPS task CALIB on the phase
calibrator to determine the phase correction as a function of time for
each individual IF band.  In the self-calibration mode, we either
selected a single strong maser line or averaged multiple maser lines
in narrow ranges of both velocity and space to perform phase
calibration. The typical solution interval was 100 seconds. After the
above calibrations, we discarded the phase solutions and the maser
data in the time intervals within which the solutions appeared to be
randomly scattered in time or had adjacent reference phases exceeding
$\sim$50$^\circ$. The phase solutions were then interpolated and
applied to all the maser data. In all phase-referencing observations
for NGC 6323, we performed an extra step of self-calibration after the
initial phase-referencing calibration, because the phase-referencing
calibration alone did not give adequate phase calibration, and
significant sidelobes were still visible after 
CLEAN deconvolution.
 
The calibrated data from multiple tracks of a
single source were combined for imaging.  We 
Fourier transformed the gridded $(u,v)$ data to make images of the
masers in all spectral channels of the IF bands that showed maser
lines, and we deconvolved the images using CLEAN
with a weighting scheme that optimized the position accuracy
(e.g. ROBUST = 0 in the task IMAGR). We fitted the detected maser
spots with elliptical Gaussians to obtain
the positions and flux densities of individual maser components.  In
Table 3 we show a representative dataset that includes the
velocities, positions, and peak intensities of the maser spots in NGC
6264. The data for all galaxies are available in the electronic version
of this paper. Note that all galaxies except NGC 4388 and
NGC 2273 have at least 2 tracks of data. Therefore, the peak flux in
Table 3 is the flux for the averaged dataset.  The actual fluxes
in an individual epoch can be higher or lower because of variability
of the masers.
 
\subsection{Relativistic Velocity Assignment} 
The sub-parsec megamaser disks presented in this paper are in 
a deep gravitational potential and the majority have recession
velocities over 1\% of the speed of light $c$. For this reason, we
made both special and general relativistic corrections to the maser
velocities in the data before we used the data to analyze the BH
masses. We made the relativistic corrections in the following way:
 
The observed Local Standard of Rest velocities $V_{\rm
op}$ listed in Table 3 are based on the ``optical'' velocity           
convention 
\begin{equation} \label{optdopeq}
{V_{\rm op} \over c} = \biggl({\nu_0 - \nu
\over \nu}\biggr)~, 
\end{equation}                                
where $\nu$ is the observed frequency and $\nu_{0}$ = 22.23508 GHz, 
the rest frequency of the H$_2$O 6$_{16}-5_{23}$ transition.      
 
Because of gravitational time dilation, the emitting frequency 
$\nu_{0}$ of a maser at distance $r$ from a compact object of mass 
$M$ and the actual observed frequency $\nu_\infty$ for an observer at 
$r = \infty$  differ by a factor 
\begin{equation} 
{\nu_0 \over \nu_\infty} = \biggl(1  + {G M \over r c^2}\biggr)~. 
\end{equation} 
For a maser in a circular orbit moving at speed $\beta_{\rm m}c$, 
balancing the gravitational and centripetal accelerations gives 
$GM/r^2 = (\beta_{\rm m} c)^2/r$, so 
\begin{equation} 
{\nu_0 \over \nu_\infty} = 1 + \beta_{\rm m}^2~. 
\end{equation} 
We multiplied the observed frequency $\nu$ of each maser line by 
$(1 + \beta_{\rm m}^2)$ to correct for gravitational time dilation. 
 
The megamaser lines at the systemic velocities were also corrected
for a transverse Doppler shift                                          
\begin{equation} 
{\nu_0 \over \nu} = 1 + \beta_{\rm m}^2/2~. 
\end{equation} 
 
The relativistically correct Doppler equation for a source moving 
radially away with velocity $v = \beta c$ is 
\begin{equation}\label{reldopeq} 
{\nu \over \nu_0} = \biggl({1 - \beta \over 1 + \beta}\biggr)^{1/2}~. 
\end{equation} 

After the observed frequency had been corrected for the gravitational
time dilation and transverse Doppler shift, we used
Equation~\ref{reldopeq} to convert the corrected frequency to its
relativistically correct radial velocity $v = \beta c$.  For the
megamaser disks, typically $\beta_{\rm m} < 0.003$.  So, for galaxies
in the Hubble flow ($z > 0.01$), the general relativistic corrections
are smaller than the special relativistic corrections.  For example,
for masers in UGC 3789, which has an optical-LSR recession velocity of
3262 km~s$^{-1}$, the special relativistic corrections range from 10
to 26 km~s$^{-1}$ whereas the general relativistic corrections range
from 0 to $\sim$ 2 km~s$^{-1}$.
 
Finally, when fitting the rotation curves, we used the relativistic formula 
for the addition of velocities to decompose the observed $\beta$ value of each 
maser spot into a common $\beta_{\rm g}$
associated with the radial velocity of the center of each megamaser
disk and an individual $\beta_{\rm m}$ associated with orbital motion
of a specific maser spot. The redshifted maser spots have $\beta_{\rm m} > 0$ and
blueshifted maser spots have $\beta_{\rm m} < 0$.
\begin{equation} 
\beta = {\beta_{\rm g} + \beta_{\rm m} \over 
1 + \beta_{\rm g} \beta_{\rm m}}~. 
\end{equation}

\section{RESULTS} 
\subsection{VLBI Images, Rotation Curves, and BH Masses}
Figure 1 shows the GBT single-dish spectra for all megamaser galaxies presented
here except for UGC 3789, which
can be found in \citet{rbg09}.  Figures 2, 3, and 4 show the VLBI maps and
the position-velocity ($P-V$) diagrams along with the fitted rotation
curves of the maser spectral components (spots)
in UGC~3789, NGC~1194, NGC~2273, NGC~2960 (Mrk~1419), NGC~4388,
NGC~6264, and NGC~6323. \citet{rbg09} published the UGC 3789 VLBI map
and the rotation curve, and we performed a new analysis of the BH mass
for this galaxy based on those data. We show the VLBI map and
rotation curve for this galaxy again for direct comparison with the other six
megamaser disks. The data points in the VLBI maps and rotation curves
are color-coded to indicate redshifted, blueshifted, and systemic
masers, where the ``systemic'' masers refer to the maser spectral
components having velocities close to the systemic velocity of the
galaxy.  Except for NGC 4388, the maser spot distributions are plotted relative to the average position of the systemic
masers.  Systemic masers are not detected in NGC 4388, so we plotted its
maser distribution relative to the dynamical center determined from
fitting the data in the $P-V$ diagram with a Keplerian rotation curve.
 
To estimate the inclination and dynamical center (i.e. the position of the BH) of each disk, we
 rotated the coordinate system to make the disk horizontal and used the fitted horizontal
line that passes through the high velocity masers as the zero point of
the y-coordinate of the dynamical center (see Figures 3 \& 4). The zero
point of the x-coordinate is defined to be the unweighted average $\theta_{\rm
x}$ of the systemic masers. 

We assumed that the systemic masers have about the same orbital radii
as the high-velocity masers to estimate the maser disk inclination
cos$^{-1}$($\langle \theta_{\rm y}^{(sys)} \rangle / \theta_{\rm r}$),
where $\langle \theta_{\rm y}^{(sys)} \rangle$ is the average
$\theta_{\rm y}$ position of the systemic masers and $\theta_{\rm r}$
is the orbital radius of the systemic masers. In principle, one can
determine $\theta_{\rm r}$ exactly only when good rotation curves for
both systemic and high-velocity masers can be obtained, and precise
acceleration measurements for the systemic masers are available. Among
our data we only have such information for UGC 3789 at this point
\citep{bra10}, and so we use $\langle$$\theta_{\rm x}$$\rangle$ of
high velocity masers as an estimate of $\theta_{\rm r}$. We note that
all our megamaser disks with systemic masers detected have
inclinations larger than 80$^{\circ}$, so assuming the disk is exactly
edge-on will only cause errors less than 1\% in the derived BH
masses. In NGC 4388 we could not measure a disk inclination, but even
if the disk were 20\arcdeg\ from edge-on, the contribution to the error
in the BH mass would be only 12\%, comparable to the 11\% error caused
by the distance uncertainty.
 
We determined the rotation curve for each megamaser disk as a function of the ``impact parameter'' defined as the projected radial offset $\theta
= (\theta_{\rm x}^2 +
\theta_{\rm y}^2)^{1/2}$ of the maser spots so that we can account for the warped 
structures in some megamaser disks. We then performed a nonlinear least-squares 
fit of a Keplerian rotation curve to the position-velocity diagram with the assumption that the
high-velocity masers lie exactly on the mid-line of the disk. In
addition, the systemic velocity of each galaxy was fitted as a free
parameter, and we report the best fits of the systemic velocities
of our megamaser galaxies in Table 1.                                    
 
The fitted Keplerian rotation curves can be expressed with the following
form:\\                                                                 
\begin{equation} 
\vert v_{\rm K} \vert = v_1 \biggl({\theta \over 1 {\rm ~mas}}\biggr)^{-1/2},
\end{equation} 
where $\vert v_{\rm K} \vert$ is the orbital velocity (after relativistic corrections) 
of the high velocity masers  and $v_{1}$ is the orbital velocity at a radius
1 mas from the dynamical center.                                        
The BH mass is  \begin{equation} 
M_\bullet = \biggl({\vert v_{\rm K} \vert^2 \theta \over G}\biggr)
D_{\rm A} ~ = \biggl({ \pi v_{1}^2 \over 6.48 \times 10^8 G}\biggr)  D_{\rm A} ~,              
\end{equation} 
where $D_{A}$ is the angular diameter distance to the galaxy.
We show all the measured BH masses in Table 4. 
 
\subsection{The Error Budget For the BH Mass}
There are three major sources of error for our BH mass calculation,
and the largest comes from the distance uncertainty. Except for NGC 4388,
we used the
Hubble distances relative to the CMB for a
cosmological model with $H_0$ = 73 km~s$^{-1}$~Mpc$^{-1}$,
$\Omega_{\rm matter}= 0.27$ and $\Omega_\Lambda = 0.73$.  Since $H_0$
is uncertain by 8~km~s$^{-1}$~Mpc$^{-1}$ \citep{fw01}, the galaxy
distances will have an rms uncertainty of 11\%.  NGC 4388 is in the Virgo
cluster and has a large peculiar velocity, so we used its Tully-Fisher
distance of 19 Mpc \citep{rus02}.  Since the distance modulus used by
\citet{rus02} is calibrated relative to the Cepheid distances provided
by the Hubble Key Project \citep{fw01} and the systematic error of the
Cepheid distances is the primary source of the Hubble constant
uncertainty, we also adopted 11\% as the rms distance error for NGC 4388.
 
The second source of BH mass uncertainty comes from fitting the
rotation curves of the masers. In our Keplerian fitting, the rms
$\sigma_\theta$  of the observed position offsets from Keplerian rotation
is usually larger than the rms position uncertainties
of the data by a factor of 1.5 to 4.  This extra scatter can lead to a
systematically different position of the dynamical center, and hence a slightly
different BH mass if we allow dynamical center position to be a free
parameter in the fit.  This excess may indicate
that we underestimated our observational errors or it may 
indicate genuine deviations from our simple model, possibly caused by
peculiar motions of the masers (e.g. perturbations from spiral density
waves) or high-velocity masers not lying precisely on the mid-line of
the disk \citep{hmp08}. If the latter is the main cause of deviation,
we estimate that the majority of the maser spots fall within $\approx
7^\circ - 10^\circ$ from the midline of the disk, depending on the
galaxy.  
  
In addition to the deviations of the masers from the midline of the maser disk, there is also an error
in the BH mass from the uncertainty of the position of the BH in the fitting. To estimate this error, we relax our assumption on the position of the BH, and allow it to be a free parameter in the fitting. 
However, since the recession velocity of the galaxy (v$_0$) and the position of the BH are correlated 
in this case, we impose a constraint on the possible positions of the BH such that the fitted
v$_0$ does not exceed v$_0$ determined from other meothods (e.g. HI measurements) beyond their error bars. Including all possible errors mentioned above, we estimate the total fitting error in BH mass to be 1\%$-$5\% depending on the galaxy.
 
The third source of error is from the absolute position errors of our
megamasers. Source position errors introduce an extra phase difference
between maser spots having different frequencies (velocities). The
relative position errors scale with velocity offset from the reference
maser feature \citep{argon07, gmr93}, which may differ from source to
source. Since the BH mass is proportional to the size of the maser
orbits, relative position shifts among masers introduce errors in the
BH mass measurements.  Among our megamasers, UGC~3789, NGC~2960, and
NGC~ 2273 have the largest absolute position errors (10 mas; see Note
(h) in Table 1). The resultant BH mass errors are $3-5$\%, whereas
this error is smaller than 0.3\% in the other megamaser galaxies.
Please note that the error for the BH mass discussed in the following
subsections does not include the distance uncertainty. Only source
position errors and errors from Keplerian fitting are evaluated, which
is the best approach for comparing our results with BH masses derived
from other techniques as the same distance can be used for comparison.
As the distance uncertainties shrink in the future, so will the
uncertainties in our BH masses.

\subsection{Notes On Individual Galaxies} 
\subsubsection{NGC~1194} 
NGC 1194 hosts a Seyfert 1.9 nucleus and has a distance of $\approx52$
Mpc.  The position angle of the maser disk (Figure 2) is $157^\circ$
and the inclination is $\approx$ 85$^\circ$. NGC 1194 has the
largest maser disk among the megamasers presented here, with an inner
and outer radius of 0.53 and 1.3 pc, respectively. The blueshifted and
redshifted masers do not appear to fall exactly on a straight line on
the sky, but they are consistent with a slightly bent thin disk. The
rotation curve (Figure 3) is consistent with Keplerian rotation, and
the measured BH mass is $(6.5\pm0.3) \times 10^7 M_\odot$, the
largest among the BH masses studied here.
 
\subsubsection{NGC~2273} 
NGC 2273 is a Seyfert 2 galaxy at a distance of $\approx26$ Mpc. The
position angle of the maser disk (Figure 2) is 153$^\circ$ with an
inclination of $\approx84^\circ$. The inner and outer radii of the
disk are 0.09 and 0.26 pc, respectively. The maser distribution shows
a hint of a warped disk. The rotation curve (Figure 3) is also
consistent with Keplerian rotation to within the errors. The BH mass
we obtain is $(7.6\pm0.4) \times 10^{6} M_\odot$.
 
\subsubsection{UGC~3789} 
UGC 3789 is a Seyfert 2 galaxy at a distance of $\approx50$ Mpc
\citep{rbg09,bra10}.  Its edge-on maser disk lies at  position
angle  $41^\circ$ East of North (Fig.~2). Although the
high-velocity masers appear to trace a flat disk, the misalignment of
the systemic masers with respect to the disk and the fact that the
systemic masers have at least two distinct centripetal accelerations
\citep{bra10} suggest that the disk is probably warped along the line
of sight. The inner and outer radii of the disk are 0.085 and 0.32 pc,
respectively. Fitting the $P-V$ diagram (Fig.~3) with a Keplerian
rotation curve gives a BH mass of $(1.12\pm0.05)\times10^7
M_\odot$. Note that this mass is nearly the same as the mass given
in \citet{bra10}, $(1.09\pm0.05)\times 10^{7} M_{\odot}$.  The very
small difference comes from choosing different positions as the
dynamical center, the uncertainty of which has been taken into account
in our estimate of the BH mass error.

\subsubsection{NGC~2960 (Mrk~1419)} 
NGC 2960 hosts a LINER nucleus at a distance of $\approx71$
Mpc. The position angle of the maser disk (Fig.~2) is
$-131^{\circ}$ and the inclination is $\approx 89^{\circ}$. The
outer parts of the disk show some warping. The inner and outer radii
of the disk are 0.13 and 0.55 pc, respectively. The rms scatter of the
high velocity masers normal to the disk is $\sigma_\perp =
71\,\mu$arcsec, which is 1.6 times larger than the rms uncertainty of
the data, so it is likely that either we underestimated the
observational uncertainties because of larger tropospheric delay errors
for low declination sources, or the thickness of the disk may not be
negligible in this megamaser. In this galaxy, we measure a BH mass of $(1.14\pm0.05) \times
10^{7} M_{\odot}$.
 
\subsubsection{NGC~4388} 
NGC 4388 is in the Virgo cluster and we adopted a Tully-Fisher
distance of 19 Mpc; it is
the nearest of the seven galaxies presented here. It hosts a
Seyfert 2 nucleus and the maser disk (Figure 2) has a position angle
of 107$^{\circ}$.  We cannot measure the inclination directly because
no systemic masers were detected. As with the other
megamasers, we assumed a flat, edge-on disk.
 
Since no systemic masers were detected, we determined the position of
the dynamical center by allowing it to be a free parameter when
fitting the high-velocity masers with a Keplerian rotation curve
(Fig.~4).  In addition, we fixed the systemic velocity of the galaxy
using an HI measurement \citep{lu03}. The resulting BH mass is
$(8.5\pm0.2)\times 10^6 M_{\odot}$. With only five maser spots
mapped, there is not sufficient data to demonstrate Keplerian
rotation or even to show that the masers
actually lie on a disk. We argue that it is likely they do because the
radio continuum jet shown by \citet{kuk95} is nearly perpendicular to
the line joining the blueshifted and redshifted masers.  However,
because of these uncertainties, the BH mass for NGC 4388 should be
used with some caution until better data are obtained.
 
\subsubsection{NGC 6264} 
At $\approx136$ Mpc, NGC 6264 (Figure 2) is the most distant object in
our current sample. The disk is slightly warped and appears to have
some thickness, but given the uncertainties in the position
measurement, our observation is consistent with a thin disk. The disk
has a position angle of $-85^{\circ}$ and an inclination of $\approx
90^{\circ}$. The inner and outer radii of the disk are 0.18 and 0.77
pc, respectively. In the $P-V$ diagram (Fig.~4), the high velocity
masers beautifully trace the Keplerian rotation curve, and we obtain a
BH mass of $(2.84\pm0.04) \times10^7 M_{\odot}$.
 
\subsubsection{NGC 6323} 
NGC 6323 is a Seyfert 2 galaxy at a distance of $\approx105$ Mpc. The
VLBI image (Fig.~2) shows a remarkably thin disk at a position angle
of $10^{\circ}$ and inclination of $\approx89^{\circ}$. The disk is
apparently warped, and the inner and outer radii are 0.13 and 0.3 pc,
respectively. The rotation curve of the high-velocity masers (Figure
4) is Keplerian, and we obtain a BH mass of $(9.3\pm0.1)\times10^{6}
M_\odot$.

\subsection{Search For Continuum Emission}
We searched for continuum emission from the vicinity of the supermassive BH
(i.e., near the systemic masers) in our megamaser galaxies by averaging
the line-free spectral channels in our data and imaging with natural
weighting to maximize the detection sensitivity. We detected no
continuum emission in all megamaser galaxies presented in this
paper. The channels averaged, the center velocities of the bands used
for averaging, and the continuum upper limits are listed in Table 5.

\section{A SUPERMASSIVE BLACK HOLE OR A CENTRAL CLUSTER OF STARS
OR STELLAR REMNANTS ?}                                                
 
The near-Keplerian shapes of the rotation curves of many circumnuclear
megamaser disks allows one to determine the enclosed (spherical) mass
within the innermost radius of the disk. While the enclosed masses of
the megamaser disks are very likely dominated by supermassive BHs, as
we assumed in the previous sections, it is still important to find
ways to justify this assumption. The question of
whether the enclosed mass can be attributed to a point mass or a
compact cluster of stars or stellar remnants has been addressed by
\citet{mao95} \& \citet{mao98}.  The main argument is that if the
lifetime of a central cluster, limited by evaporation or 
collision timescales, is significantly shorter than the age of its host
galaxy, then it is unlikely to persist, and a central supermassive BH
would be required to account for the enclosed mass. Here, the
evaporation timescale t$_{\rm evap} \approx 136t_{\rm relax}$ ($t_{\rm
  relax}$ is the half-mass relaxation timescale; Binney \& Tremaine 2008)
is considered to be the upper limit of the lifetime of any bound stellar system
whereas the collision timescale is the characteristic timescale that a
star suffers a physical collision (i.e. an inelastic encounter; Binney \& Tremaine 2008).
 
To estimate the lifetime of the central cluster, we follow
\citet{mao95} and assume that the central cluster has the Plummer
density distribution \citep{plu15}:
\begin{equation}\label{plummereq} 
\rho(r) = \rho_0 \biggl(1 + {r^2\over r_{\rm c}^2}\biggr)^{-5/2}~, 
\end{equation} 
where $\rho_0$ is the central density and $r_{\rm c}$ is the core radius.
(The reason for choosing the Plummer distribution is described in section 2.1 of \citet{mao98})

We constrain $\rho_0$ and $r_{\rm c}$ by fitting the position-velocity
diagram of the megamaser disk with the rotation curve of a Plummer
cluster:                                                               
\begin{equation} 
v_{\rm P} = \biggl[ {G M_\infty r^2 \over 
(r_{\rm c}^2 + r^2)^{3/2}}\biggr]^{1/2}~, 
\end{equation} 
where $M_{\infty} = 4 \pi \rho_0 r_c^3/3$ is the total mass of the
cluster. Here, $M_{\infty}$ and $r_{\rm c}$ are fitted as free
parameters. From these two parameters, we calculated
$\rho_{0}=3M_{\infty}/4\pi r_c^3$. In all cases M$_\infty$ is very close to the ``enclosed'' mass measured from the Keplerian rotation
curve fit in Section 3, and the differences are less than $4-18\%$. 
In Figures 3 \& 4, we show the fitted Plummer rotation curves along with 
the fitted Keplerian rotation curves.

Rather than using the method described above to constrain the core
radius r$_{c}$, in some cases we could apply another approach that
places even tighter constraints.  We note that the Plummer rotation
curve does not decrease monotonically with radius; instead the
rotation curve turns over at a maximum rotation speed $v_{\rm max}$:
\begin{equation}\label{vmaxeq} 
v_{\rm max} = \biggl({2 G M_\infty \over 3^{3/2} r_{\rm c}}\biggr)^{1/2}.
\end{equation} 
Having a maximum rotation velocity is not unique to a Plummer cluster. 
It is a general feature for clusters having the same form of density profile with the exponent smaller than $-3/2$. 

The core radius $r_{\rm c}$ of a Plummer sphere having 
maximum rotation speed $v_{\rm max}$ is 
\begin{equation}\label{rmaxeq} 
r_{\rm c}^{(\rm max)} = {2 G M_\infty \over 3^{3/2} v_{\rm max}^2}~. 
\end{equation} 
We used Equation 12 to estimate r$_{c}$ for the megamaser disks in NGC
1194, NGC 2273, NGC 2960 and NGC 4388.  In these cases, we do not have
very well-sampled or high quality rotation curves (see Figs.~3 \& 4),
so using r$_{\rm c}^{(\rm max)}$ from Equation 12 actually sets a
tighter constraint on the core radius of the cluster than using
rotation-curve fitting. For these four cases, we use the highest observed velocity in the PV diagram
as an estimate (lower limit) of $v_{max}$ and use it to calculate r$_{\rm c}^{(\rm max)}$ with
M$_\infty$ from the Plummer rotation curve fitting from Equation 10. For UGC 3789, NGC
6323, and NGC 6264, we used $r_{c}$ from rotation curve fitting because
the qualities of the rotation curves are good, and they give tighter constraints on
$r_{\rm c}$. In Table 4, we give the Plummer model parameters for all of our
megamaser disks.

We constrained the lifetime of the Plummer cluster $\emph{T}_{\rm age}$
by first requiring the cluster not evaporate in a timescale less than
the age of its host galaxy ($\geq 10$ Gyr if the galaxy has formed
before $z=2$). This requirement sets an upper limit to the mass of the
constituent stars of the cluster because evaporation is unimportant so
long as the mass of its stars satisfies the following
equation:

\begin{equation}\label{evapeq} 
\biggl( { m_\star \over M_\sun}\biggr) \lesssim 
[\ln(\lambda M_\infty/m_\star)]^{-1} 
\biggl({r_{\rm h} \over 0.01\,{\rm pc}}\biggr)^{3/2} 
 \biggl( {M_\infty \over 10^8 M_\sun}\biggr)^{1/2}~, 
\end{equation} 
where $m_\star$ is the mass of each star, $\lambda \approx 0.1$ 
\citep{bin08}, and $r_{\rm h}$ is the radius of half total mass ($r_{\rm
h} \approx 1.305 r_{\rm c}$). We call the maximum m$_{\star}$ that satisfies the above equation
m$_{\rm max}$ (Table 4) and used it to calculate the collision timescale 
of the Plummer cluster with the Plummer model 
parameters: 
\begin{equation}\label{tcolleq} 
t_{\rm coll} = \biggl[16 \sqrt{\pi} n_\star \sigma_\star r_\star^2 
\biggl( 1 + {G m_{max} \over 2 \sigma_\star^2 r_\star}\biggr)\biggr]^{-1}~,
\end{equation} 
where $n_\star=\rho_{0} / m_{\rm max}$ is the number density of stars,
$\sigma_\star$ is the rms velocity dispersion of the stars, and
$r_\star$ is the stellar radius \citep{mao98, bin08}. If t$_{\rm
coll}$ $<$ 10 Gyr, then $\emph{T}_{\rm age}$ is
constrained by t$_{\rm coll}$ and we can rule out the Plummer cluster as
an alternative to the BH.  If t$_{\rm coll}$ $\geq$ 10 Gyr, we cannot
rule out a cluster whether $\emph{T}_{\rm age}$ is dominated by
evaporation or collision. %%% I don't understand the grammar of this sentence.

For UGC~3789 we obtained an upper limit $m_{max} \approx 0.12 M_\sun$
to the mass of individual stars and a lower limit $N \gtrsim 1.0
\times 10^8$ to the number of stars in the cluster. The mass limit
directly rules out neutron stars as the constituents of the cluster,
and only brown dwarfs, very-low mass stars, or white dwarfs are
possible. If the constituents of the cluster are brown dwarfs, 
the collision timescale for our Plummer model in UGC~3789 is
$t_{\rm coll} < 5.0 \times 10^6$ years, much less than the age of a
galaxy. The timescale is even shorter if the constituent stars are
main-sequence stars. However, if the cluster is composed of white
dwarfs, the collision timescale can be as long as $2.7 \times 10^7$
years, but this is still much shorter than the lifetime of a galaxy.
Therefore, we conclude that a compact cluster is not likely to survive
long, and the dominant mass at the center of UGC~3789 is a
supermassive BH. For all other megamaser galaxies except NGC 1194, the
lifetimes of the central clusters are also shsorter than the age of a
galaxy (Table 4). The constraints on lifetime
weakly rule out massive clusters in NGC 4388 and NGC~6264
($\emph{T}_{\rm age} \approx 7 \times 10^{9}$ yr) but strongly rule
out clusters in the others ($\emph{T}_{\rm age} < 10^{9}$ yr).
 
In summary, by setting  tight constraints on the sizes and central mass
densities of possible Plummer clusters, we have been able to strongly rule
out clusters as the dominant central masses in UGC~3079, NGC~
2273, NGC~6323, and NGC~2960 and weakly rule out the clusters in NGC~
4388 and NGC~6264. We argue that supermassive black holes are the
dominant masses in these megamasers.  Together with the Milky Way
Galaxy, NGC~4258, and~M31 \citep{kmd01}, the number of galaxies with
strong evidence to rule out a massive star cluster as the dominant
central mass increases from three to seven.
 
\section{Other BH mass measuring techniques} 
 
It is important to compare results from different methods for
measuring BH masses, because such comparisons can provide insight into
potential systematic errors for each method \citep{snuker09,kmd04, hum09,
gre10}. Comparing optically measured dynamical BH
masses with those from the H$_{2}$O megamaser method is especially
valuable, since the megamaser galaxies with Keplerian rotation curves
provide the most direct and accurate BH mass measurements for external galaxies, 
These maser BH masses can be used to test the more commonly used BH mass measuring
techniques in the optical, such as the stellar or gas dynamical methods. 

One cannot meaningfully compare the stellar or gas dynamical BH masses
with our maser BH masses unless the gravitational spheres of influence
can be resolved. Otherwise, even if the optically determined dynamical
masses agree with the maser BH masses within the errors, the
uncertainty can be too large to tell the real accuracy of the stellar-
or gas-dynamical method. The angular diameters of the spheres of
influence of our maser BH masses, 2$R_{\rm inf}$ (Table 4), range from
$0\,\farcs006$ to $0\,\farcs06$. Among these megamasers, only the
spheres of influence in NGC 4388 and NGC 1194 can be barely resolved
by the HST (resolution $\approx0\,\farcs07$ at $\lambda \approx 6500 \AA$)
or the VLT
\footnote{The Very Large Telescope} assisted with adaptive optics
(resolution $\approx0\,\farcs1$ at $\lambda = 2\,\mu$m). Therefore, of
the galaxies here, stellar or gas dynamical measurements are feasible
only for these two galaxies. We have obtained VLT time to measure the
BH mass in NGC 4388, and we will apply the stellar dynamical method to
this galaxy and compare the BH mass to the maser BH mass in the
future.

Another commonly used BH mass measuring technique is the virial
estimation method \citep{gh06, mk08, vo09}. We are able to compare
this technique with the megamaser disk method in four galaxies, and
this is the first time that the virial estimation method is directly
tested.  In this method, one estimates the BH mass as $M_{\rm BH} =
\emph{f}~R_{\rm BLR}\sigma_{line}^2/G$, where \emph{f} is an unknown
factor that depends on the structure, kinematics, and orientation of
the broad-line region (BLR), $\emph{R}_{\rm BLR}$ is the radius of the
BLR, $\sigma_{line}$ is the gas velocity dispersion observed in the
BLR, and G is the gravitational constant. Here, we adopt the latest
empirically determined $\langle f \rangle = 5.2\pm1.3$ from
\citet{woo10}.  \citet{ves09} suggests that the virial method is
accurate to a factor of $\approx4$.  Since we cannot directly detect
the BLRs in megamaser galaxies that have Seyfert 2 nuclei, we
estimated $\sigma_{line}$ using the scattered ``polarized broad
lines'' (PBL) from the hidden BLRs in those four megamaser galaxies
with detected PBLs: NGC 1068, NGC 4388, NGC 2273, and Circinus.  We
estimated $\emph{R}_{\rm BLR}$ via the $\emph{L}_{(2-10 kev)}-R_{\rm
  BLR}$ correlation \citep{kmn05}. Table 6 shows the resulting virial
BH masses for these four galaxies.
 
Under the assumption that the observed linewidth of broad H$_{\alpha}$
or H$_{\beta}$ emission approximates the ``intrinsic
linewidth''\footnote{By ``intrinsic linewidth'', we mean the linewidth
  one would measure if the BLRs were observed directly as in Type 1
  AGNs}, we find that for NGC~1068, NGC~ 2273, NGC~4388 and Circinus,
the BH masses at the 1 $\sigma$ level measured by the virial estimation
method agree within a factor of 5 with the megamaser BH masses.  So, the
scatter of the virial BH masses relative to the maser BH
masses is comparable to the factor of 4 accuracy expected for the
virial method. It is possible that the slightly larger difference between 
the maser and viral BH masses as compared with the expected accuracy for 
the virial method is partly caused by the larger intrinsic scatter in the 
$\emph{L}_{(2-10 kev)}-R_{\rm BLR}$ correlation which we used here than that 
in the more generally used $\lambda L_{\lambda}(5100\AA)-\emph{R}_{\rm BLR}$ 
correlation (e.g. Kaspi et al. 2005) in the optical.
 
The application of the virial method to megamaser galaxies is
described in detail in the Appendix.

%\section{Discussion} 

\section{Maser BH masses and the $M-\sigma_{\star}$ relation} 
 
Table 4 shows the BH masses for all seven megamaser galaxies. 
Along with the previously published BH masses from megamaser observations, 
it is interesting to see that the maser
BH masses, except in Circinus, are within a factor of 3 of
$2.2\times10^{7} M_{\odot}$, with the uncertainty of each BH mass
$\leq$ 12\%. This small range of masses could be a selection
bias. Disk megamasers are preferentially detected in Seyfert 2 spiral
galaxies, and since the local active BH mass function for Seyfert 2
galaxies peaks at $M_{\rm BH}\approx 3\times10^{7} M_{\odot}$
\citep{hec04}, we may be limited to BHs in this range of masses by
analyzing disk megamasers.

Our new maser BH masses more than double the number of galaxies having
dynamical BH masses $M_{\rm BH}\sim10^{7} M_{\odot}$.  These
measurements play a particularly important role in constraining the
$M-\sigma_{\star}$ relation at the low-mass end of known nuclear BH
masses.  In a companion paper, \citet{gre10} find that the maser
galaxies as a group fall significantly below the $M-\sigma_{\star}$
relation defined by more massive elliptical galaxies.  As a result,
the $M-\sigma_{\star}$ relation that fits later-type and lower-mass
galaxies has both a larger scatter and lower zeropoint than the
relation for elliptical galaxies alone. However, there is a potential
caveat that measuring robust (well-defined) $\sigma_{\star}$ in
late-type galaxies is challenging \citep{gre10}, and the contribution
of systematic biases in $\sigma_{\star}$ to the deviation from the
$M-\sigma_{\star}$ relation still needs to be explored in the
future. With this caveat in mind, the observed deviations from the
$M-\sigma_{\star}$ relation at low mass imply that the relation may
not be a single, low-scatter power law as originally proposed, and our
BH mass distribution is consistent with the best simulations of galaxy growth 
regulated by ``radio'' feedback; see Figure~4 of Croton et al. (2006).

In addition to our seven galaxies with maser BH masses, we currently have VLBI
datasets for another four disk megamaser candidates, and we are
monitoring the spectra of more than six disk megamaser candidates that
are currently too faint to be observed with VLBI, but may flare up in
the future. Along with these galaxies and more megamasers we may
discover in the future with the GBT, we expect to obtain a larger
sample of maser BH masses which could help to clarify the
$M_{\rm BH}$-$\sigma_{\star}$ relation.
 
\section{Summary} 
Our main conclusions are the following: 
 
\begin{itemize} 
\item[1.] The maser distributions in all seven megamaser galaxies are consistent
with edge-on circumnuclear disks surrounding  central massive
objects in the active galactic nuclei. The inner radii of the disks
are between 0.09 and 0.5\,pc, similar to all previously published
megamaser disks. The rotation curves of all seven megamaser disks are
consistent with Keplerian rotation.  Four of the megamaser disks reveal
evidence for warps.
                                             
\item[2.]  VLBI observations of circumnuclear megamaser disks are the only means to measure directly 
the enclosed mass and the mass density well within the radius of the gravitational
sphere of influence of the central mass in these galaxies.  The high central mass densities (0.12 to 60 $\times 10^{10}
M_\sun$~pc$^{-3}$) (within the central ~0.3 pc) of the seven megamaser disks indicate that in all, except one, 
disks, the central mass is dominated by a supermassive BH
rather than an extremely dense cluster of stars or stellar remnants.    
                 
\item[3.] The BH masses measured are all within a factor of 3 of
$2.2\times10^{7} M_{\odot}$ and the accuracy of each BH mass is
primarily limited by the accuracy of the Hubble constant, currently $\pm11$\%. 
The narrow range of BH mass distribution may reflect selection from 
the local active-galaxy mass function.  
                 
\item[4.] Stellar dynamics cannot be applied to the majority of the
  megamasers presented here to measure BH masses with high
  precision. The gravitational spheres of influence in all cases
  except NGC 1194 and NGC 4388 are too small to be resolved by current
  optical telescopes. Observations with the VLT for measuring the BH mass
  in NGC 4388 with stellar-dynamical modeling are in progress.
                                             
\item[5.] Under the assumption that the broad emission linewidths can
  be estimated from polarized scattered light, we have calibrated for the first
  time the BH mass determination by the virial estimation method based on 
  optical observations.   With the latest empirically determined $\langle f \rangle = 5.2\pm1.3,$
  the virial estimated BH mass is within a factor of 5 at the 1 $\sigma$ level 
  of the accurate BH mass based on megamaser disks in NGC 1068, NGC 2273, NGC 4388, and Circinus.
  This is comparable to the factor of 4 accuracy expected for the virial
  estimation method.
 
\item[6.] The accurate BH masses in the seven megamaser galaxies
contribute to the observational basis of the $M-\sigma_{\star}$ relation
at the low-mass end. The deviation from the mean relation of the several
accurate maser BH masses suggests that the relation may not be a single, 
low-scatter power law as originally proposed, which has interesting implications
for the universality of the $M-\sigma_{\star}$ relation \citet{gre10}.                                
 
\end{itemize}

\acknowledgements The National Radio Astronomy Observatory is a
facility of the National Science Foundation operated under cooperative
agreement by Associated Universities, Inc. We thank Ed Fomalont for
his kind help with our VLBI data reduction and Lincoln Greenhill for
his substantial assistance for this project. C.Y. Kuo thanks Mark
Whittle for his numerous insightful comments on our project. This
research has made use of NASA's Astrophysics Data System Bibliographic
Services, and the NASA/IPAC Extragalactic Database (NED) which is
operated by the Jet Propulsion Laboratory, California Institute of
Technology, under contract with the National Aeronautics and Space
Administration.
 
\appendix 
\section{Applying The Virial Estimation Method
to Megamaser Galaxies}                                                  
%%%, based on relatively simple spectral measurements 
The virial estimation method for measuring BH masses in AGNs
(e.g. \citet{gh06}, \citet{mk08}, \citet{vo09}) uses the broad-line
region (BLR) gas as a dynamical tracer. It is usually only applied to
Type 1 AGNs, where the BLRs can be directly observed. In this method,
one estimates the BH mass with the following equation:
\begin{equation} 
{M_\bullet} = {f R_{\rm BLR}\sigma_{\rm line}^2 \over G}~, 
\end{equation} 
where $f$, $\emph{R}_{\rm BLR}$, and $\sigma_{\rm line}$ have been
defined in section 5. Since one can directly detect light from the
BLRs in Type 1 AGNs, $\sigma_{\rm line}$ can be measured from the broad
line spectra and one can use the continnum luminosity $\lambda L_{\lambda}(5100\AA)$
to estimate $\emph{R}_{\rm BLR}$ via the $\lambda L_{\lambda}(5100\AA)-\emph{R}_{\rm BLR}$
correlation (e.g. \citet{kmn00, kmn05}).                                       
 
In a Type 2 AGN, including the megamaser
galaxies we study here, our line-of-sight to the BLR is blocked by
heavy dust extinction, so one cannot directly measure $\lambda
L_{\lambda}(5100\AA)$ and $\sigma_{\rm line}$. Instead, one can probe
the BLRs in megamaser galaxies with polarized scattered light and hard
X-rays. Among all the megamaser galaxies with measured BH masses,
polarized scattered light from the BLRs has been detected in NGC 1068,
NGC 4388, NGC 2273, and Circinus (see Table 6), and X-ray measurements
are also available for these four galaxies.
 
We can estimate $\sigma_{\rm line}$ from the relation $\sigma_{\rm
line}/V_{\rm FWHM}=2.09\pm0.45$ from \citet{woo10}, where V$_{\rm
FWHM}$ is the FWHM width of polarized scattered light in the broad
H$_{\alpha}$ line (NGC 2273, NGC 4388, and Circinus) or H$_{\beta}$ line (NGC
1068). The major concern with these linewidths is that the observed
values may not be the same as the linewidth one would measure if the
BLRs could be observed directly. We identify two effects that can
induce such a difference from the well-studied case NGC 1068
\citep{mgm91}. First, the broad H$_{\beta}$ lines in the polarized
flux spectra can contain a contribution from the narrow H$_{\beta}$
lines. Without removing the contribution from narrow lines, the BLR
linewidth may be underestimated by $\sim$20-30\% in NGC 1068. Second,
the polarized emission from AGNs may originate from light being
scattered by electrons with a temperature a few times $10^{5}$ K,
which results in significant thermal broadening ($\sim$50\%) of the
spectral lines. Because of these two effects, there will be systematic
errors in $\sigma_{\rm line}$ if one directly uses the observed
linewidth of the polarized lines to estimate $\sigma_{\rm line}$.
Among the four megamaser galaxies we consider here, only NGC 1068 has
been studied in enough detail to remove these two effects. Luckily,
the two effects change the linewidth in opposite directions, and hence
could offset each other to a certain extent. As in NGC 1068, if no
correction is made for these two effects, the systematic error will be
only $\sim$10\%, just slightly larger than the measurement
error. Without knowing the actual contributions of these two effects
for the other three galaxies, we assume that the two effects cancel
each other to the same extent as in NGC 1068 and use the observed
linewidths as the approximations for the intrinsic widths. The reader
should be aware of this caveat when interpreting the comparisons made.
 
In this work, $\emph{R}_{\rm BLR}$ was estimated from the ``intrinsic''
$\emph{L}_{\rm (2-10 kev)}$ of the nuclear region via the
$\emph{L}_{\rm (2-10 kev)}-\emph{R}_{\rm BLR}$ correlation
\citep{kmn05}. Since three (NGC 1068, NGC 2273, and Circinus) of
the four megamaser galaxies considered here are Compton-thick
(i.e. the X-ray absorbing column density is $>$ 10$^{24}$ cm$^{-2}$),
we paid particular attention to how the $\emph{L}_{\rm (2-10 kev)}$
were measured. The Compton thick nature of these AGNs is a problem
because the intrinsic radiation is mostly suppressed and the
X-ray spectrum is dominated by the reflected or scattered
components. It is difficult to measure the actual absorbing column
density and it is very likely that the intrinsic hard X-ray luminosity
is severely underestimated, e.g. \citet{lhk06, bdm99}. Therefore, we
excluded those measurements that did not consider the Compton-thick
nature of these sources and failed to give the absorbing column density
in the expected range. We mainly considered those measurements from
either data with appropriate modeling or from observations with
instruments capable of directly measuring the transmission components
of X-ray above 10 kev. We took at least two different measurements for
each galaxy from the literature and used the average value to
calculate R$_{BLR}$ from the correlation in
\citet{kmn05}.
 
Given $R_{BLR}$ and $\sigma_{\rm line}$, we estimated the BH masses
using Equation A1 with the empirically determined $\langle f \rangle =
5.2^{+1.3}_{-1.3}$ from \citet{woo10}. The resultant BH masses were
compared with the maser BH masses in section 5.

\appendix

\newpage 
%table 1 : Basic information of the maser galaxies 
\begin{deluxetable}{lccccrccll} 
\tablewidth{0 pt} 
\tablecaption{The Megamaser Sample} 
\tablehead{ 
\colhead{}               & \colhead{R.A.}       & 
\colhead{Decl.}          & \colhead{$\delta$RA} & 
\colhead{$\delta$DEC}    &\colhead{V$_{\rm sys}$}    & 
\colhead{$\delta$V$_{\rm sys}$}  & 
\colhead{Spectral}       & \colhead{Hubble}     \\ 
\colhead{Name}           & \colhead{(J2000)}    & 
\colhead{(J2000)}        & \colhead{(mas)}      & 
\colhead{(mas)}          & \colhead{(km~s$^{-1}$)} & 
\colhead{(km~s$^{-1}$)}    & 
\colhead{Type}           & \colhead{Type} 
} 
\startdata                                 
NGC 1194   & 03:03:49.10864\tablenotemark{a} & $-$01:06:13.4743\tablenotemark{a}
& 0.2 & 0.4 & 4051 & 15 &Sy 1.9  &  SA0+   \\                          
NGC 2273   & 06:50:08.65620\tablenotemark{b} &  60:50:44.8979\tablenotemark{b}
& 10\tablenotemark{h}  & 10\tablenotemark{h}   & 1832     & 15
&Sy 2    &  SB(r)a   \\     
UGC 3789   & 07:19:30.9490\tablenotemark{c}  &  59:21:18.3150\tablenotemark{c}
& 10\tablenotemark{h} & 10\tablenotemark{h}    & 3262   & 15
&Sy 2    & (R)SA(r)ab \\                                                                        
NGC 2960   & 09:40:36.38370\tablenotemark{d} &  03:34:37.2915\tablenotemark{d}
& 10\tablenotemark{h}  & 10\tablenotemark{h}   & 4945     & 15
&Liner   & Sa?       \\                                                 
NGC 4388   & 12:25:46.77914\tablenotemark{e} &  12:39:43.7516\tablenotemark{e}
& 0.4  & 0.3   & 2527      & 1     & Sy 2   &  SA(s)b   \\   
NGC 6264   & 16:57:16.12780\tablenotemark{f} &  27:50:58.5774\tablenotemark{f}
& 0.3   & 0.5   & 10213    & 15    &Sy 2    &  S ?      \\              
NGC 6323   & 17:13:18.03991\tablenotemark{g} &  43:46:56.7465\tablenotemark{g}
& 0.2   & 0.4    & 7848     & 10    &Sy 2    &  Sab      \\            
\enddata 
\tablecomments{(1)The systemic (recessional) velocities of the galaxies, V$_{\rm sys}$,
listed here are based on the ``optical'' velocity convention (i.e. no relativistic corrections are made)
, measured with respect to the Local Standard of Rest (LSR). Except NGC 4388, the
systemic velocities V$_{\rm sys}$ are obtained from fitting a
Keplerian rotation curve to the observed data as described in section
3. The uncertainties $\delta$V$_{\rm sys}$ given here include both the
fitting error and a conservative estimate of the systematic error. The
fitting error is typically only about 5 km~s$^{-1}$ and the systematic
error is from possible deviation of the position of the BH
from ($\theta_x,\theta_y)=(0,0)$ (see section 3), which is assumed in
our rotation curve fitting. For NGC 4388, we adopt the observed HI
velocity from \citet{lu03}, which has a small but perhaps unrealistic
error. (2) The positions of UGC 3789, NGC 1194, NGC 6323, and NGC 4388
refer to the location of maser emission determined from our VLBI
phase-referencing observations. The positions of NGC 2273 and NGC 2960
are determined from K-band VLA A-array observations of continuum
emission from program AB1230 and maser emission from AB1090,
respectively.  The maser position for NGC 6264 is derived from
a phase-referencing observation in the VLBA archival data (project
BK114A).  (3) The Seyfert types and morphological classifications are
from NASA/IPAC Extragalactic Database (NED).  }
\tablenotetext{a}{The position of the maser spot at V$_{\rm op} =$ 4684
kms$^{-1}$, where V$_{\rm op}$ is the ``optical'' velocity of the maser spot relative to the LSR.}                                                            
\tablenotetext{b}{The position of the radio continuum emission observed
at 21867.7 MHz and 21898.9 MHz. }  
\tablenotetext{c}{The position of the maser spot at V$_{\rm op} =$ 2689
kms$^{-1}$.} 
\tablenotetext{d}{The position of the maser spot at V$_{\rm op} =$ 4476
kms$^{-1}$.}                                                            
\tablenotetext{e}{The position of the maser spot at V$_{\rm op} =$ 2892
kms$^{-1}$.}                                         
\tablenotetext{f}{The average position of the masers with velocities
from V$_{\rm op} =$ 10180 kms$^{-1}$ to V$_{\rm op} =$ 10214 kms$^{-1}$}      
\tablenotetext{g}{The position of the maser spot at V$_{\rm op} =$ 7861
kms$^{-1}$.}                                                            
                                                          
\tablenotetext{h}{For all positions derived from VLA observations, we use 10\,mas as the
actual position error, rather than use the fitted error from the VLA
data, which are only a few mas for these galaxies. The reason is that
the systematic error caused by the imperfect tropospheric model of the
VLA correlator can be as large as a few to 10s\,mas. A 10\,mas
position error usually leads to a $\approx 3-5$\,\% error in the BH
mass for the megamasers presented here. Note that although the
position for UGC 3789 is derived from a VLBI phase-referencing
observation with a phase calibrator $2.1^{\circ}$ away, the position
of this calibrator is derived from a VLA observation. So, we also use
10 mas as the actual position accuracy for UGC 3789.}
                                     
\end{deluxetable}

\begin{deluxetable}{lccllcc} 
\tablewidth{0 pt} 
\tablecaption{Observing Parameters} 
\tablehead{ 
\colhead{Experiment}               & \colhead{}         & 
\colhead{}          & \colhead{}             & 
\colhead{Synthesized Beam}        & \colhead{Sensitivity}             & 
\colhead{Observing}         \\ 
\colhead{Code}           & \colhead{Date}      & 
\colhead{Galaxy}        & \colhead{Antennas\tablenotemark{a}}   
&                                                                       
\colhead{(mas x mas,deg)\tablenotemark{b}}  & \colhead{(mJy)}      & 
\colhead{Mode\tablenotemark{c}}             } 
\startdata 
BB261    & 2005 Mar 06  & UGC 3789 & VLBA, GB, EB  & 0.55$\times$0.55,
39.0  &  1.0  & Self-cal. \\                                            
BB242B   & 2007 Nov 13  & NGC 1194 & VLBA, GB      & 2.45$\times$0.36,$-$7.5
& $\sim1.5$  & Self-cal.\\                                                    
BB242D   & 2008 Jan 21  & NGC 1194 & VLBA, GB      & 1.83$\times$0.44,
$-$10.5& 2.1  & Phase-ref. \\                                             
BB261B   & 2009 Feb 28  & NGC 2273 & VLBA, GB      & 0.69$\times$0.39,
$-$22.4& 0.5  & Self-cal. \\                                              
BB261F   & 2009 Apr 06  & NGC 6264 & VLBA, GB, EB  & 0.93$\times$0.38,$-$24.8&
0.5    & Self-cal. \\                                                   
BB261H   & 2009 Apr 18  & NGC 6264 & VLBA, GB, EB  & 1.02$\times$0.29,$-$14.7&0.3
& Self-cal. \\                                                          
BB231E   & 2007 Apr 07  & NGC 6323 & VLBA, GB, EB  & 0.58$\times$0.24,$-$13.6&
0.8   & Phase-ref.    \\                                                
BB231F   & 2007 Apr 08  & NGC 6323 & VLBA, GB, EB  & 0.58$\times$0.27,$-$14.8&
0.8   & Phase-ref.    \\                                                
BB231G   & 2007 Apr 09  & NGC 6323 & VLBA, GB, EB  & 0.92$\times$0.32,$-$5.8
& 0.8  & Phase-ref.    \\                                               
BB231H   & 2007 Apr 29  & NGC 6323 & VLBA, GB, EB  & 0.90$\times$0.28,$-$13.5
& 1.3 & Phase-ref.    \\                                                
BB242F   & 2008 Apr 13  & NGC 6323 & VLBA, GB      & 0.95$\times$0.44,$-$9.4
& 0.8  & Phase-ref.    \\                                               
BB242E   & 2008 Apr 14  & NGC 6323 & VLBA, GB      & 0.90$\times$0.49,$-$9.2
& 0.8  & Phase-ref.    \\                                               
BB242G   & 2008 Apr 16  & NGC 6323 & VLBA, GB      & 0.79$\times$0.45,$-$10.2
& 1.0    & Phase-ref.    \\                                             
BB242H   & 2008 Apr 19  & NGC 6323 & VLBA, GB      & 0.80$\times$0.45,$-$4.2&
0.3 & Self-cal.    \\                                                   
BB242J   & 2008 May 23  & NGC 6323 & VLBA, GB, EB  & 0.58$\times$0.21,$-$8.1
& 1.7   & Phase-ref.    \\                                              
BB242M   & 2009 Jan 11  & NGC 6323 & VLBA, GB, EB  & 0.52$\times$0.29,$-$17.8
& 0.4 & Self-cal.    \\                                                 
BB242R   & 2009 Apr 17  & NGC 6323 & VLBA, GB, EB  & 0.87$\times$0.45,12.2
& 0.5   & Self-cal.    \\                                               
BB242S   & 2009 Apr 19  & NGC 6323 & VLBA, GB, EB  & 0.57$\times$0.26,$-$21.2
& 0.4   & Self-cal.    \\                                               
BB242T   & 2009 Apr 25  & NGC 6323 & VLBA, GB, EB  & 0.75$\times$0.23,$-$19.0
& 0.8   & Self-cal.    \\                                               
BB248    & 2009 Mar 07  & NGC 2960 & VLBA, GB, EB  & 0.97$\times$0.45,
5.3 & 2.5      & Self-cal.  \\                                          
BB261C   & 2009 Mar 20  & NGC 2960 & VLBA, GB, EB  & 1.11$\times$0.48,2.6
& 1.1  & Self-cal.    \\                                                
BB261D   & 2009 Mar 23  & NGC 2960 & VLBA, GB, EB  & 1.26$\times$0.44,
$-$2.1 & 1.3   & Self-cal.    \\                                          
BB184C   & 2006 Mar 26  & NGC 4388 & VLBA, GB, EB  & 1.23$\times$0.32,
$-$9.27 & 1.8   & Phase-ref.   \\                                         
 
\enddata 
\tablenotetext{a}{VLBA: Very Long Baseline Array; GB: The Green Bank
Telescope of NRAO; EB: Max-Planck-Institut f\"{u}r Radioastronomie
100 m antenna in Effelsberg, Germany.}                                  
\tablenotetext{b}{Except for program BB184C, this column shows the
average FWHM beam size and position
angle (PA; measured east of north) at the frequency of systemic masers.
For BB184C, the FWHM and PA is measured at the frequency of red-shifted
masers because no systemic maser is detected (FWHM and PA differ
slightly at different frequencies). }                                   
\tablenotetext{c}{``Self-cal.'' means that the observation was conducted
in the ``self-calibration'' mode and ``Phase-ref.'' means that we used the
``phase-referencing'' mode of observation.}                                                   
\end{deluxetable} 
 
%Table 3: Maser velocities and positions 
\begin{deluxetable}{lrrrrrr} 
\tablewidth{0 pt} 
\tablecaption{Sample data for NGC 6264} 
\tablehead{ 
\colhead{V$_{\rm op}$\tablenotemark{a}} & \colhead{RA\tablenotemark{b}}  
& \colhead{$\delta$RA\tablenotemark{b}}  & \colhead{Decl.\tablenotemark{b}}   
& \colhead{$\delta$DEC\tablenotemark{b}}  & \colhead{F$_\nu$\tablenotemark{c}}  & \colhead{$\sigma_F$\tablenotemark{c}} 
\\                                                                      
\colhead{(km~s$^{-1}$)}  &\colhead{(mas)}  
&\colhead{(mas)}         &\colhead{(mas)}         
& \colhead{(mas)}  & \colhead{(mJy/B)} & \colhead{(mJy/B)}        } 
\startdata 
10918.33  &   0.399  &  0.008  &  $-$0.021  &  0.018  &   6.0  &   0.3 \\
10914.71  &   0.407  &  0.010  &  $-$0.001  &  0.021  &   4.8  &   0.3\\
10911.09  &   0.401  &  0.005  &   0.000  &  0.011  &  10.2  &   0.4\\
10907.47  &   0.397  &  0.013  &   0.041  &  0.029  &   3.7  &   0.3\\
10903.85  &   0.395  &  0.011  &  $-$0.050  &  0.027  &   3.8  &   0.4\\
10900.23  &   0.388  &  0.009  &   0.044  &  0.022  &   4.6  &   0.3\\
10885.74  &   0.369  &  0.019  &   0.053  &  0.038  &   2.6  &   0.4\\
10871.26  &   0.400  &  0.018  &   0.061  &  0.037  &   2.9  &   0.3\\
10856.77  &   0.467  &  0.011  &  $-$0.038  &  0.029  &   3.8  &   0.3\\
10853.15  &   0.484  &  0.005  &  $-$0.038  &  0.011  &   9.7  &   0.4\\
10849.53  &   0.490  &  0.005  &  $-$0.006  &  0.011  &   9.4  &   0.3\\
10845.91  &   0.492  &  0.002  &  $-$0.016  &  0.006  &  18.3  &   0.4\\
10842.29  &   0.494  &  0.002  &  $-$0.005  &  0.005  &  24.4  &   0.4\\
10838.67  &   0.505  &  0.004  &  $-$0.010  &  0.009  &  11.7  &   0.4\\
10835.05  &   0.543  &  0.012  &  $-$0.094  &  0.032  &   3.5  &   0.4\\
10831.43  &   0.542  &  0.017  &  $-$0.039  &  0.031  &   3.1  &   0.3\\
10827.81  &   0.518  &  0.012  &  $-$0.101  &  0.027  &   3.9  &   0.3\\
10824.19  &   0.533  &  0.007  &  $-$0.040  &  0.016  &   6.9  &   0.4\\
10820.56  &   0.493  &  0.012  &  $-$0.041  &  0.026  &   3.8  &   0.3\\

\enddata 
\tablecomments{Sample of data for NGC 6264. The entirety of data
for all galaxies is available in the electronic version. }
\tablenotetext{a}{Velocity referenced to the LSR and using the optical definition (no relativistic corrections).}  
\tablenotetext{b}{East-west and north-south position offsets and
uncertainties measured relative to the average position of the
systemic masers in the VLBI map (Figure 2). Position uncertainties reflect fitted random errors
only. In NGC~1194 and NGC~2960, there may be additional uncertainties caused
by the poorer tropospheric delay calibrations due to their low
declinations (low elevations during the observations).}
\tablenotetext{c}{Fitted peak intensity and its uncertainty in mJy~beam$^{-1}$. }    
\end{deluxetable}

%table 4: Basic information of the maser galaxies 
% For the dynamical timescale, only show the longest one.******** 
% Check Lincoln's paper and see whether any important properties
                                                  
\begin{deluxetable}{lcccccccccc} 
\tablewidth{0 pt} 
\tablecaption{The BH Masses and Basic Properties of the Maser Disks} 
\tablehead{ 
\colhead{}
&  \colhead{Dist.} & \colhead{BH mass} & \colhead{Disk Size} &
\colhead{P.A.} & \colhead{Incl.}  & \colhead{r$_{\rm c}$}  &
\colhead{$\rho_{0}$} & \colhead{$m_{\rm max}$} &                                                                       
\colhead{$\emph{T}_{\rm age}$} &\colhead{$R_{\rm inf}$}  \\ 
\colhead{Name}   & \colhead{(Mpc)}   & \colhead{(10$^{7}$
M$_{\odot}$)}  & \colhead{($\rm pc$)}      & \colhead{($^\circ$)} &
\colhead{($^\circ$)}  & \colhead{(pc)} &
\colhead{(M$_{\odot}~pc^{-3}$)} & \colhead{(M$_{\odot}$)} &                                                                       
\colhead{(yr)}     & \colhead{(arcsec)}   } 
\startdata 
          
NGC 1194 &  52  & 6.4$\pm$0.3   & 0.53-1.30 & 157  & 85  
& 0.250 & 1.2$\times 10^{9}$  & 12 & $>$1.0$\times$10$^{10}$ & 0.033\\              
NGC 2273 &  26  & 0.76$\pm$0.04 & 0.028-0.085 & 153  & 84  
& 0.015 & 6.0$\times 10^{11}$ & 0.05   & $<$2.2$\times$10$^{6}$ & 0.010 \\  
UGC 3789 &  50  & 1.12$\pm$0.05 & 0.09-0.32 &  41  & $>$ 88
& 0.024 & 2.0$\times 10^{11}$ & 0.12   & $<$2.7$\times$10$^{7}$ &0.010   \\                  
NGC 2960 &  71  & 1.14$\pm$0.05 & 0.13-0.36 &$-$131  & 89  
& 0.055 & 1.8$\times 10^{10}$ & 0.5   & $<$9.3$\times$10$^{8}$ & 0.005 \\            
NGC 4388 &  19  & 0.84$\pm$0.02 & 0.24-0.29 & 107  & --   
& 0.090 & 3.3$\times 10^{9}$  & 0.9 & $<$6.6$\times$10$^{9}$ & 0.034 \\ 
NGC 6264 & 136  & 2.84$\pm$0.04 & 0.23-0.78 & $-$85  & 90  
& 0.083 & 1.3$\times 10^{10}$ & 1.3  & $<$8.0$\times$10$^{9}$ & 0.012  \\            
NGC 6323 & 105  & 0.93$\pm$0.01 & 0.13-0.30 &  10  & 89  
& 0.046 & 2.3$\times 10^{10}$ & 0.3   & $<$4.7$\times$10$^{8}$ & 0.003 \\    
             
\enddata 
\tablecomments{Col(1): Galaxy name; Col(2): Distances we adopt
from NED; Col(3): Black hole masses measured
in our project. The mass uncertainty here only includes errors caused
by source position uncertainty and from fitting a Keplerian rotation
curve for a given distance. The actual uncertainty is limited by the error
of the Hubble Constant H$_{0}$ (11\%); Col(4): Sizes of the inner and outer edge of
the maser disks; Col(5): Position angle (PA) of the disk plane measured
east of north. PA equals zero when the blueshifted side of the disk plane has zero 
East offset and positive North offset; Col(6): Inclination 
of the maser disk. Note that the inclination of NGC 4388 is unconstrained 
because we did not detect systemic masers; Col(7): The core radius of 
the Plummer cluster in parsecs. For NGC 1194, NGC 2273, NGC 2960, 
and NGC 4388, the core radii are derived from Equation 12. For UGC 3789, 
NGC 6264, and NGC 6323, the radii are derived from the Plummer rotation 
curve fitting; Col(8): The central mass density of the Plummer cluster
$\rho_{0}=3M_{\infty}/4\pi r_c^3$. Here, $M_{\infty}$ is obtained from the Plummer rotation curve fitting
; Col(9): The maximum stellar mass of the Plummer cluster 
below which the cluster will not evaporate in less than 10 Gyr. In all cases except NGC 1194,
a cluster of neutron stars can be directly ruled out because m$_{\rm max}$ is less than $\approx$ 1.4 M$_{\odot}$;
Col(10): Lifetime ($\emph{T}_{\rm age}$) of a cluster. The values shown here are limited by 
the collision timescale, which is the maximum lifetime of the cluster composed of either main-sequence 
stars, brown dwarfs, white dwarfs, or neutrons stars; Col(11) The radius of the gravitational sphere of influence for
the maser BHs in arcsec. We calculate $R_{inf}$ using Equation 1 of \citet{barth03} with the bulge 
velocity dispersion measurements from \citet{gre10}.   }                        
 
\end{deluxetable} 
 
%In NGC 2273, we average over channels 1 through 25 of the redshifted dual-circularly polarized band centered at V$_{\op}=$ 2800 km~s$^{-1}$, and

\begin{deluxetable}{lccll} 
\tablewidth{0 pt} 
\tablecaption{Upper limit on Continnum Emission from Megamaser Galaxies} 
\tablehead{ 
\colhead{Galaxy}      & \colhead{V$_{\rm Center}$}
 & \colhead{V$_{\rm Range}$}  &                                                 
\colhead{I$_{2\sigma}$}   & \colhead{Project Code}
\\                                                                      
\colhead{}  & \colhead{(km~s$^{-1}$)}  &\colhead{(km~s$^{-1}$)}
&                                                                       
\colhead{(mJy)}        & \colhead{}         } 
\startdata 
NGC 1194  &  3568      &  3489 - 3660  &    $<$ 0.34 &  BB242B     \\
NGC 2273  &  2800  &     2868 - 2909  &  $<$ 0.18    & BB261B \\ 
NGC 2960  &  5315    &   5296 - 5354  &  $<$ 0.14    & BB261C, BB261D     \\
NGC 4388  &  2600   &  2881 - 2966     &  $<$ 0.42   & BB184C      \\
NGC 6264  &  9710     &  9726 - 9776    & $<$ 0.16   & BB261F, BB261H       \\
NGC 6323  & 8100, 7650 &   8005 - 8200, 7556 - 7750  &  $<$ 0.08   & BB242M, BB242R, BB242S, BB242T     \\

\enddata 
\tablecomments{Col(1): Name of the galaxy; Col(2):The central Optical-LSR velocity of
the bands used to search for continuum emission; Col(3) The velocity
range corresponding to the selected channels. Note that the channels
selected are chosen to be free of any maser lines except for NGC
2960. For this galaxy, we don't have line-free channels, and we
averaged the bands centered at the systemic velocity of the galaxy
and searched for continuum emission offset from the systemic maser
emission; Col(4) The $2\sigma$ detection limit of the continuum
emission; Col(5) The data used for continuum detection. }
      
\end{deluxetable}

%Table 6: Comparison between maser and virial BH mass
\begin{deluxetable}{lcccccll}
\tablewidth{0 pt}
\tablecaption{Comparison of Maser BH Mass with Mass from Virial Estimation}
\tablehead{
\colhead{Galaxy}      & \colhead{maser BH}         & \colhead{virial BH}      & \colhead{R$_{BLR}$}  & \colhead{V$_{\rm FWHM}$}   & \colhead{Reference}             & \colhead{L$_{X}$(2-10 kev)}     & \colhead{Reference}  \\
\colhead{}            & \colhead{(10$^{6}$ M$_{\odot}$)} &\colhead{(10$^{6}$ M$_{\odot}$)}   &
\colhead{(light-days)} &\colhead{(km~s$^{-1}$)}   & \colhead{}    & \colhead{(10$^{42}$ ergs~s$^{-1}$)}  & \colhead{}  }
\startdata
NGC 1068  &  8.6$\pm$0.3 &  9.0$\pm$6.6  & 4.6$\pm$3.0 & 2900  & 1  & 6.5, 2.6         & 5, 6     \\ 
NGC 4388  &  8.5$\pm$0.2 &  7.1$\pm$4.9  & 2.0$\pm$1.2 & 3900  & 2  & 0.8, 0.9,1.9,1.0 & 7,8,9,10 \\
NGC 2273  &  7.6$\pm$0.1 &  4.3$\pm$2.8  & 2.2$\pm$1.2 & 2900  & 3  & 1.0, 1.7         & 11,12 \\
Circinus  &  1.7$\pm$0.3 &  4.8$\pm$3.2  & 1.9$\pm$1.1 & 3300  & 4  & 1.1, 1.0, 1.2    & 13,14,15 \\  
\enddata
\tablecomments{Col(1): Galaxy name; Col(2) BH mass measured from the megamaser
technique. The BH mass of NGC 1068 is taken from \citet{lb03} and
Circinus from \citet{gbe03}. The BH masses of NGC 4388 and NGC 2273
are from this paper. Col(3): BH mass measured from the virial
estimation method. Col(4): The size of the Broad Line Region (BLR)
calculated using the L$_{(2-10 kev)}$$-$R$_{BLR}$ correlation
\citep{kmn05}. Col(5): The full width at half maximum (FWHM) of the
observed broad line (H$_{\beta}$ for NGC 1068 and H$_{\alpha}$ for NGC
4388, NGC 2273, and Circinus). In NGC 2273, only the full width at
zero intensity (FWZI) is given, so we estimate the FWHM ``by eye''. We
assume the measurement uncertainty is 200 km~s$^{-1}$ in all
cases. Col(6) Reference for the linewidth measurement. Col(7)
Intrinsic hard X-ray (2-10 kev) luminosity. We used the average
luminosity to calculate the size of BLR. The error of the luminosity
is taken to be the standard deviation of the adopted
luminosities. Note that the distances used to calculate the intrinsic
luminosities are 15.4, 19.0, 26.0, and 4.0 Mpc for NGC 1068, NGC 2273,
NGC 4388, and Circinus, respectively. Col.(8) Reference for the
adopted X-ray luminosities.  \\ References $-$ 1. \citet{mgm91}
2. \citet{ho97} 3. \citet{mbk00} 4. \citet{omc98} 5. \citet{lhk06}
6. \citet{obc03} 7. \citet{cpb06} 8. \citet{ag09} 9. \citet{bdm99}
10. \citet{flk99} 11. \citet{tih02} 12. \citet{ath09} 13.\citet{ywm09}
14. \citet{mgm99} 15. \citet{sw01} }
\end{deluxetable}

\newpage 
\begin{figure}[ht] 
\begin{center} 
%\vspace*{-0.3 cm} 
%\hspace*{-2 cm} 
\includegraphics[angle=0, scale=0.6]{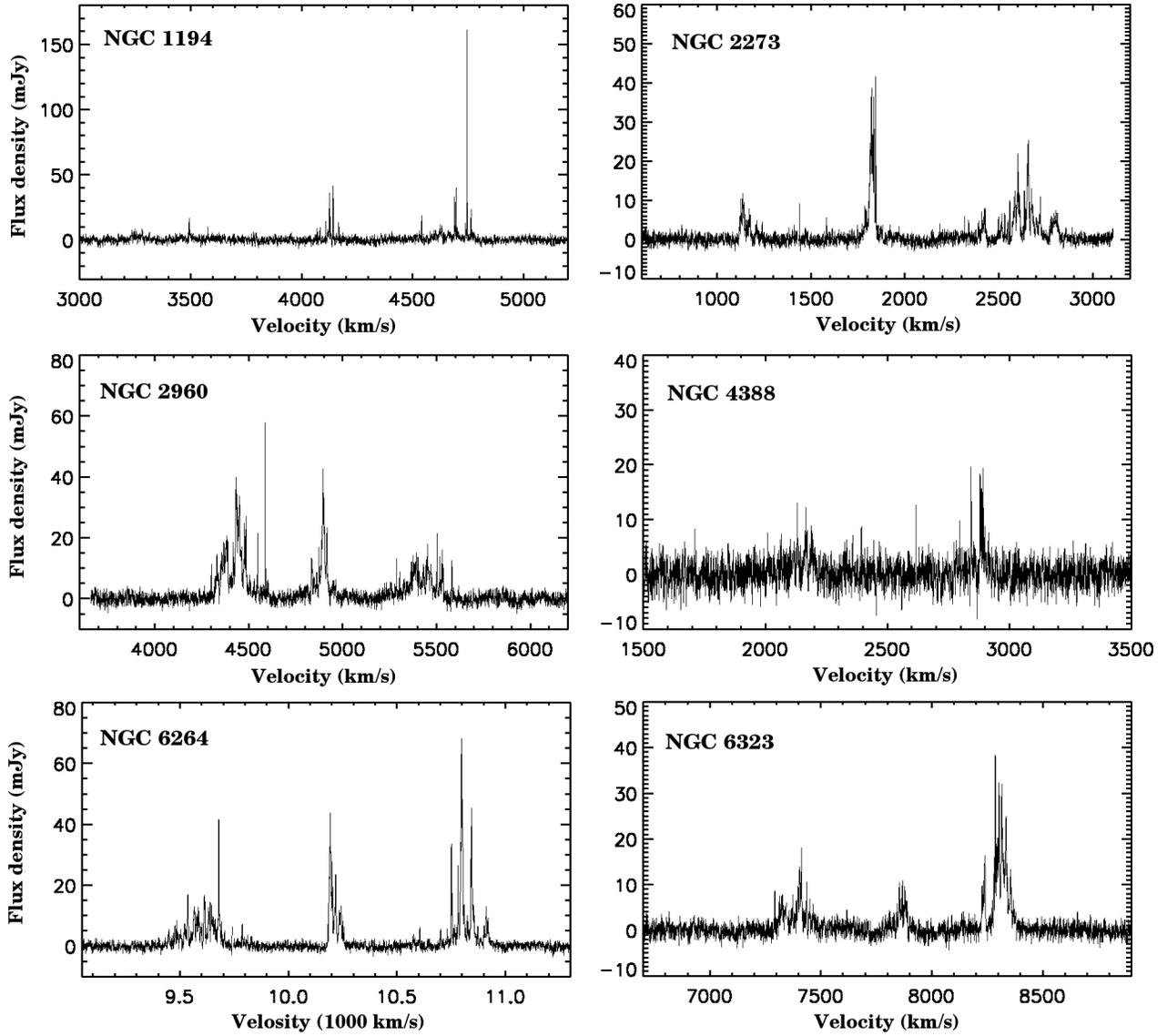}
\vspace*{0.0 cm} 
\caption{Characteristic H$_{2}$O maser spectra. The x-axis
shows LSR velocities based on optical definition. Flux densities of
masers can vary significantly, so the spectra shown here are just
representative for particular epochs: January 13 2008 for NGC 1194; February 21 2009
for NGC 2273; April 2 2009 for NGC 2960 (Mrk~1419); November 30
2005 for NGC 4388; March 31 2009 for NGC 6264; and April 6 2000 for NGC 6323.  }
\end{center} 
\end{figure}

\begin{figure}[ht] 
\begin{center} 
%\vspace*{-0.3 cm} 
%\hspace*{-2 cm} 
\includegraphics[angle=0, scale=1.0]{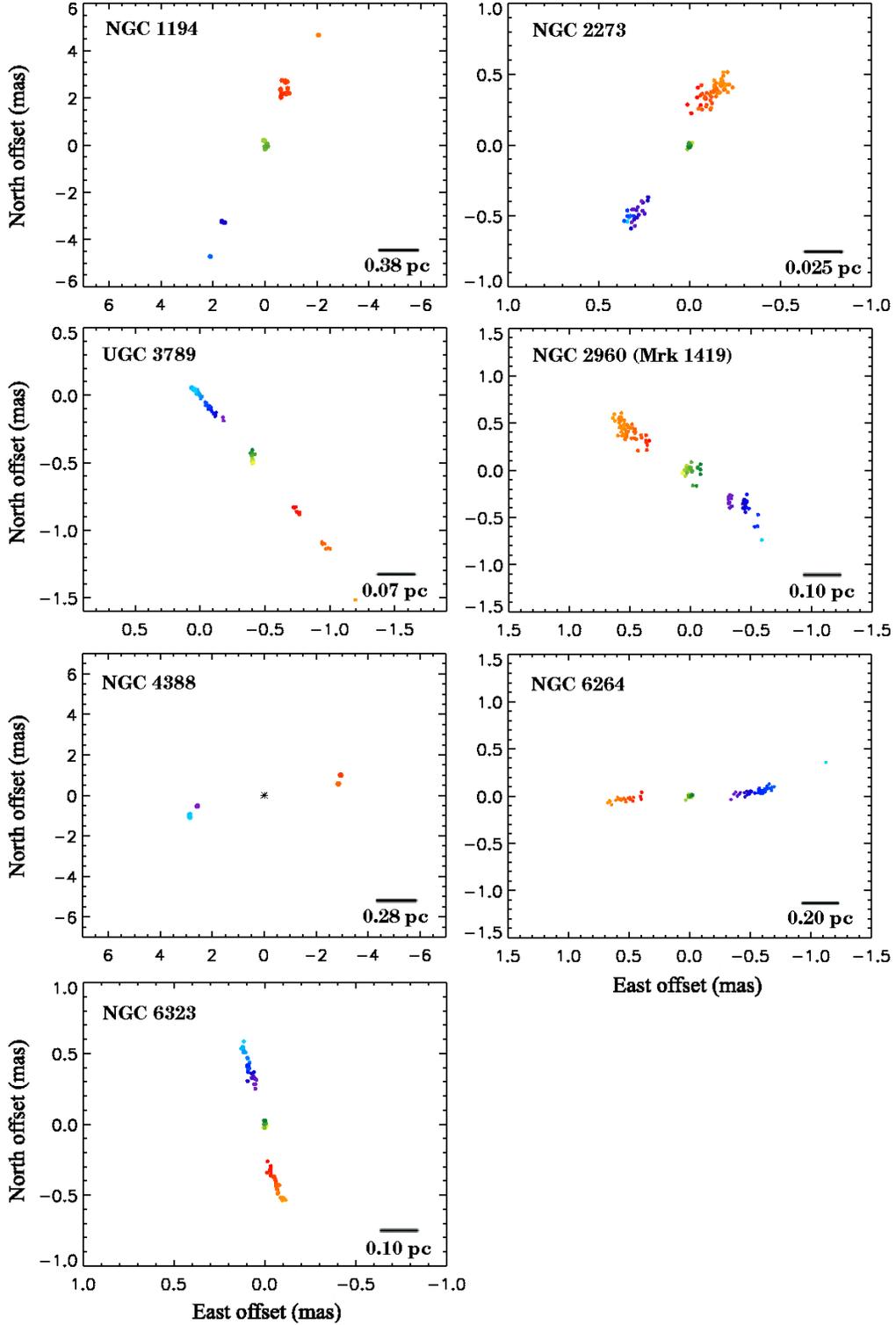} 
\vspace*{0.0 cm} 
\caption{VLBI maps for the seven 22\,GHz H$_2$O masers megamasers
analyzed. The maps are color-coded to indicate redshifted,
blueshifted, and systemic masers, where the ``systemic'' masers refer
to the maser components having recessional velocities close to the
systemic velocity of the galaxy. Except NGC 4388, maser distributions
are plotted relative to the average position of the systemic masers. For
NGC 4388, in which the systemic masers are not detected, we plot the
maser distribution relative to the dynamical center determined by
fitting the high velocity features with a Keplerian rotation curve.}
\end{center} 
\end{figure} 
 
\begin{figure}[ht] 
\begin{center} 
%\vspace*{-0.3 cm} 
%\hspace*{-2 cm} 
\includegraphics[angle=0, scale=1.0]{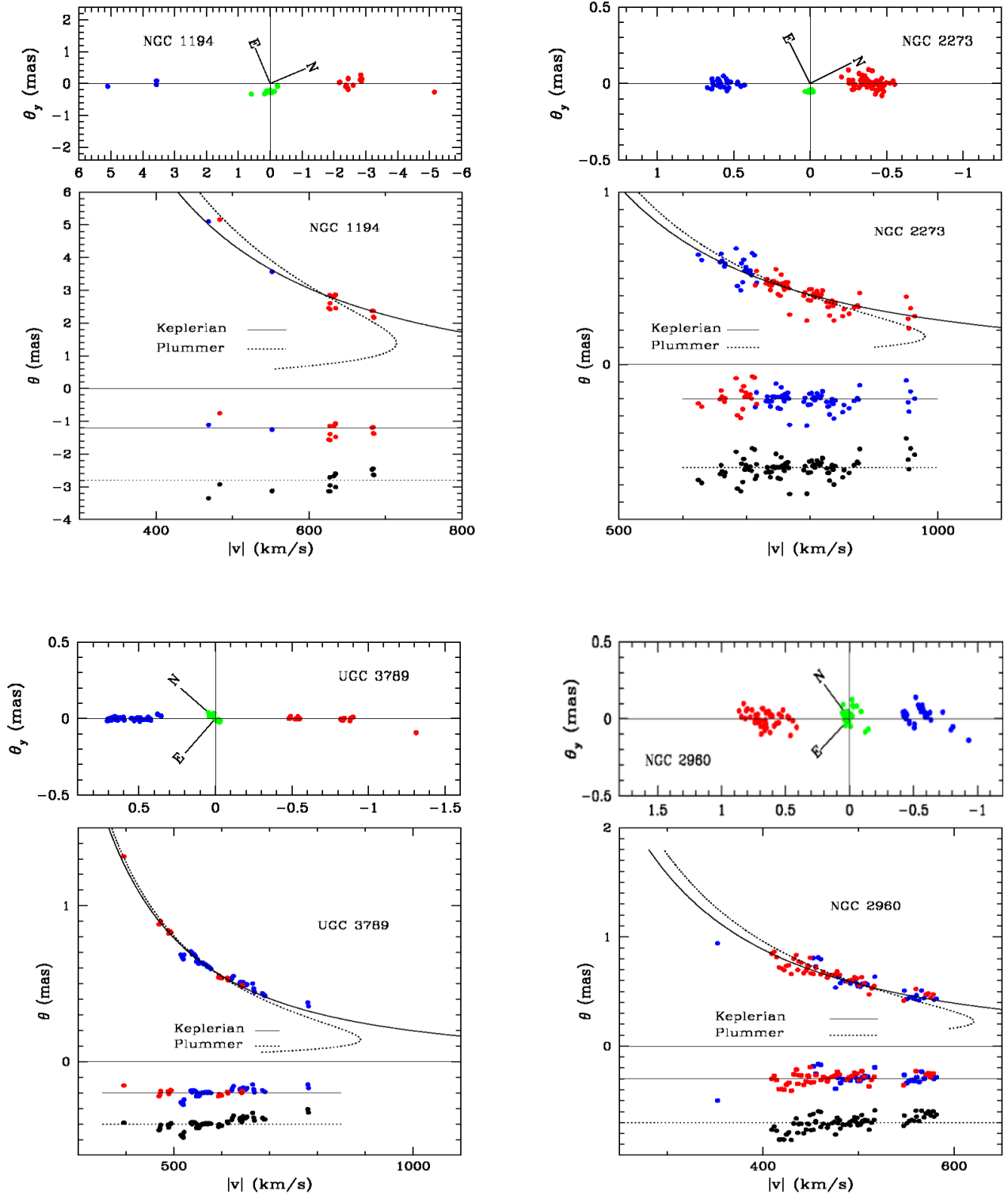} 
\vspace*{0.0 cm} 
\caption{Maser distributions (top panels) and rotation curves (bottom panels) for
NGC 1194 , NGC 2273, UGC 3789, and NGC 2960. The maser distribution
has been rotated to horizontal to show the scatter in the maser
positions and the offset of the systemic masers from the plane defined
by high-velocity masers more clearly. The coordinate system is chosen
to place the centroid of the high-velocity maser disk (blue and red
points) at $\theta_{\rm y} = 0$ and the centroid of the systemic
masers (green points) at $\theta_{\rm x} = 0$. The axes for the maps
show relative position in milliarcseconds, and North (N) and east (E)
are indicated by directional arrows on each map. The bottom panel for
each galaxy shows the rotation curves of the redshifted and
blueshifted masers (red and blue points on the curves) plotted with
the best-fit Keplerian (solid curve) and Plummer (dotted curve)
rotation curves. The velocities shown in the figure are the LSR
velocities after the special and general relativistic corrections. The
residuals (data minus Keplerian curve in red and blue; data minus
Plummer curve in black) are in the bottom part of each figure. Note
that we plot the rotation curve with the impact parameter $\theta$
(mas) as the ordinate and rotation speed $\vert v \vert$ (km s$^{-1}$)
as the abscissa for the convenience of fitting.}

\end{center} 
\end{figure} 

\begin{figure}[ht] 
\begin{center} 
%\vspace*{-0.3 cm} 
%\hspace*{-2 cm} 
\includegraphics[angle=0, scale=1.0]{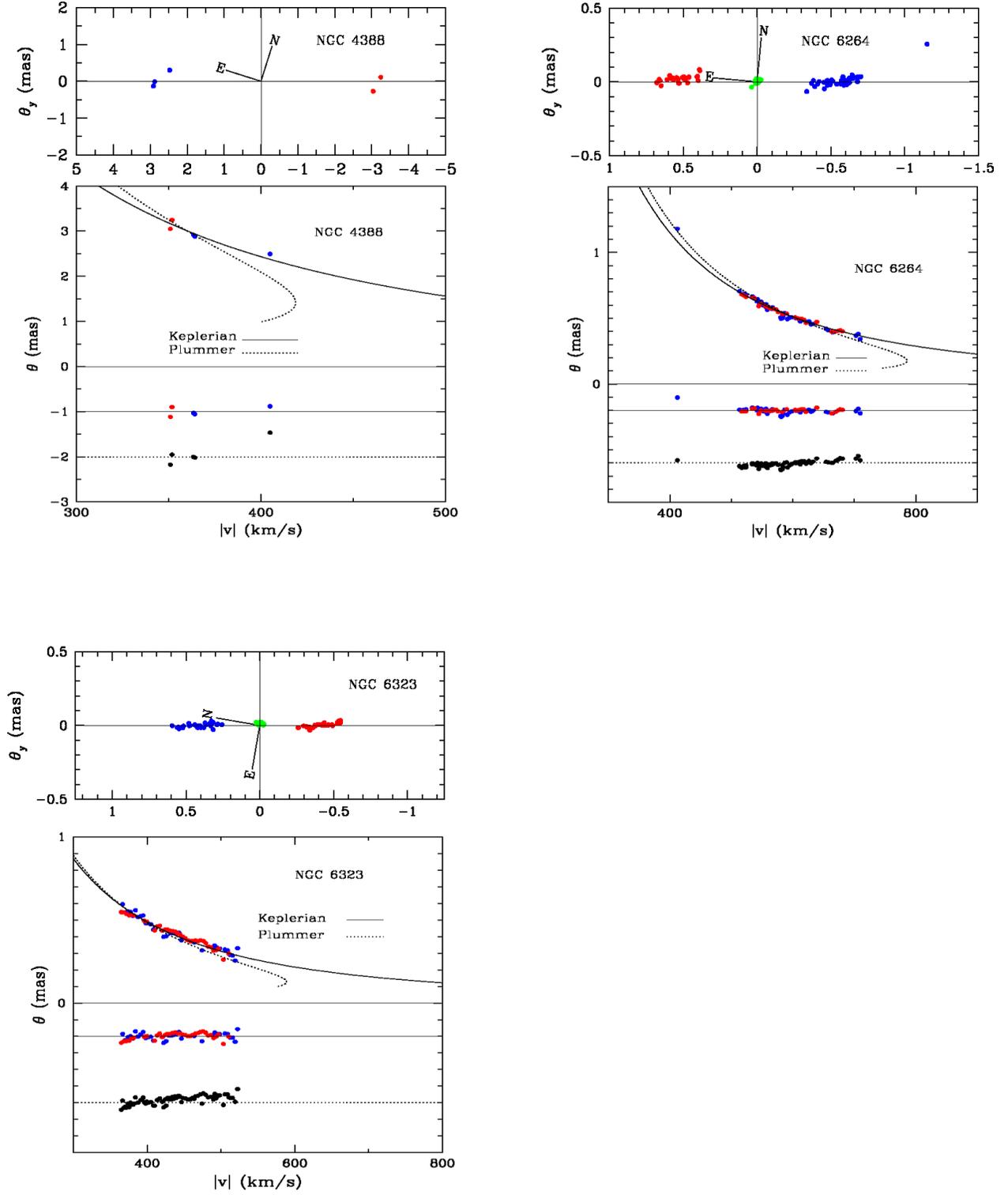} 
\vspace*{0.0 cm} 
\caption{Maser distributions (Top panel) and rotation curves (Bottom panel) for NGC 4388
, NGC 6264, and NGC 6323. Please refer to the caption of Figure 3 for the description of this figure.}                   

\end{center} 
\end{figure} 

\end{document}